\documentclass[pre,superscriptaddress,showpacs,longbibliography,reprint]{revtex4-1}

\usepackage{hyperref}

\usepackage{color}
\usepackage[usenames,dvipsnames]{xcolor}
\usepackage{amsmath,amsthm,amssymb}
\usepackage{graphicx}
\usepackage{epsfig}
\usepackage{dcolumn}
\usepackage{bm}
\usepackage{mathrsfs}
\usepackage{multirow}
\usepackage[all]{xy}
\usepackage{pbox}
\usepackage{verbatim}
\usepackage{bbold}
\usepackage{braket}
\usepackage{dsfont}
\usepackage{color,soul}

\DeclareMathOperator{\Tr}{Tr}
\def\(({\left(}
\def\)){\right)}
\def\[[{\left[}
\def\]]{\right]}

\newcommand{\be}{\begin{equation}}
\newcommand{\ee}{\end{equation}}
\newcommand{\ben}{\begin{eqnarray}}
\newcommand{\een}{\end{eqnarray}}
\newcommand{\beq}{\begin{equation}}
\newcommand{\eeq}{\end{equation}}

\newcommand{\emptyDM}{\ket{\circ}\hspace{-3pt}\bra{\circ}}

\usepackage{graphicx}

\begin{document}

\title{Numerical Simulation of Critical Dissipative Non-Equilibrium Quantum Systems with an Absorbing State}

\author{Edward Gillman}
\affiliation{School of Physics and Astronomy, University of Nottingham, Nottingham, NG7 2RD, UK}
\affiliation{Centre for the Mathematics and Theoretical Physics of Quantum Non-Equilibrium Systems,
University of Nottingham, Nottingham, NG7 2RD, UK}

\author{Federico Carollo}
\affiliation{School of Physics and Astronomy, University of Nottingham, Nottingham, NG7 2RD, UK}
\affiliation{Centre for the Mathematics and Theoretical Physics of Quantum Non-Equilibrium Systems,
University of Nottingham, Nottingham, NG7 2RD, UK}

\author{Igor Lesanovsky}
\affiliation{School of Physics and Astronomy, University of Nottingham, Nottingham, NG7 2RD, UK}
\affiliation{Centre for the Mathematics and Theoretical Physics of Quantum Non-Equilibrium Systems,
University of Nottingham, Nottingham, NG7 2RD, UK}
\date{\today}

\begin{abstract}
The simulation of out-of-equilibrium dissipative quantum many body systems is a problem of fundamental interest to a number of fields in physics, ranging from condensed matter to cosmology. For unitary systems, tensor network methods have proved successful and extending these to open systems is a natural avenue for study. In particular, an important question concerns the possibility of approximating the critical dynamics of non-equilibrium systems with tensor networks. Here, we investigate this by performing numerical simulations of a paradigmatic quantum non-equilibrium system with an absorbing state: the quantum contact process. We consider the application of matrix product states and the time-evolving block decimation algorithm to simulate the time-evolution of the quantum contact process at criticality. In the Lindblad formalism, we find that the Heisenberg picture can be used to improve the accuracy of simulations over the Schr\"{o}dinger approach, which can be understood by considering the evolution of operator-space entanglement. Furthermore, we also consider a quantum trajectories approach, which we find can reproduce the expected universal behaviour of key observables for a significantly longer time than direct simulation of the average state. These improved results provide further evidence that the universality class of the quantum contact process is not directed percolation, which is the class of the classical contact process. 
\end{abstract}

\maketitle

\section{Introduction}

Despite the recent experimental progress in probing the emergent behaviour of out-of-equilibrium ensembles of cold atoms or trapped ions, \cite{Syassen2008,Kim2010,Barreiro2011,Bohnet2016, Lienhard2018, Wade2018}, a clear understanding of these quantum non-equilibrium systems remains a major challenge. While in classical settings the study of such systems -- of their phases and critical phenomena -- are well developed, options for going beyond semi-classical treatments or the physics of exactly solvable quantum models are rather limited. This is especially true for open (dissipative) systems, which are of interest as they have potential to display a rich set of novel non-equilibrium physics -- e.g. critical dynamical behaviour or dynamical phase transitions -- not possible in closed (unitary) settings.

Here, we explore the simulation of critical dynamics in dissipative quantum many body systems, in the case of the quantum contact process (QCP) \cite{Marcuzzi2016,Buchhold2017,Roscher2018,Minjae2019,Carollo2019}. The QCP is an attractive model to study for a number of reasons: Firstly, the QCP is the coherent version of the well-understood classical contact process (CCP) \cite{Hinrichsen2000}, which exhibits a non-equilibrium phase transition (NEPT) in the directed percolation (DP) universality class, even in $1d$. Formulating both the CCP and QCP in the Lindblad formalism then allows for a direct comparison between the performance of a given numerical approach in the classical and quantum cases. 

Secondly, the QCP contains an absorbing state, which has been suggested to make simulations of quantum systems more challenging \cite{Carollo2019}. This idea can be explored by comparing the simulation of dynamics performed in the Sch\"{o}dinger picture with that of the Heisenberg picture: In the contact process, as in other dissipative systems, there is an asymmetry between dynamics in the Schr\"{o}dinger picture and Heisenberg picture. In particular, in the Heisenberg picture the absorbing state is absent. It is then of interest to investigate any difference in performance between simulations in the two. Finally, the critical physics of the $1d$ QCP is similar to that of the CCP, in the sense that key observables display power-law behaviour but with different exponents. This means that, at criticality, different numerical methods or approaches to the dynamics can be compared by their ability to reproduce the expected power-laws. Furthermore, since the universality class of the QCP is currently debated, \cite{Marcuzzi2016,Roscher2018,Carollo2019}, it is of considerable interest in its own right to make estimates of critical exponents, comparing these with previous estimates and known cases.  

To simulate the non-equilibrium dynamics of the QCP, we apply matrix product states (MPSs) and the time-evolving block-decimation (TEBD) algorithm \cite{Vidal2004, Schollwock2011, Paeckel2019}. This algorithm belongs to a more general class of tensor network (TN) methods, well established for the simulation of closed quantum systems in $1d$, which have also been applied to dissipative quantum systems previously in a number of cases \cite{Bonnes2014,Jaschke2018,Cui2015,Mascarenhas2015,Gangat2017,Kshetrimayum2017,Cuevas2013,Cuevas2016,Werner2016}. 
In the context of studying dissipative quantum dynamics, a key question for TN methods is whether different approaches, such as quantum trajectories (QTs) as opposed to the Lindblad master equation, can lead to substantially different accuracies. This question has been explored previously in \cite{Bonnes2014,Jaschke2018} and it has been suggested that in high-entanglement scenarios QTs might prove more accurate.

In the case of simulating the critical QCP with TEBD, we find the following key results: First, in the Lindblad formalism, the Heisenberg picture can be used to improve the accuracy of simulations beyond that of the Schr\"{o}dinger picture and this can be explained by considering the evolution of operator-space entanglement entropy. Second, we find that a QTs approach leads to a significant improvement in the approximation of key universal observables, i.e., the reproduction of power-law behaviour for longer times. Finally, using the results from QTs, we show that the estimated exponents of the QCP lie far from those of DP, providing further evidence that the QCP belongs to a universality class different to DP.

While we focus on the application of TEBD to the critical dynamics of the QCP, we note there are also a number of other methods available with which it would be interesting to compare results. For example, cluster mean-field \cite{Jin2016}, variational minimisation \cite{Weimer2015} and algorithms based on time-dependent variational Monte Carlo with neural networks have all shown promise for the simulation of open quantum systems \cite{Carleo2017,Nagy2019,Vicentini2019,Hartmann2019,Yoshioka2019}.

The layout of the paper is as follows: In Section \ref{sect:model} we discuss the classical and quantum contact processes. In Section \ref{sect:method} we provide a brief overview of the TEBD algorithm and MPSs for the unfamiliar reader. Section \ref{sect:simulation_double_space} then shows the results for the simulation of the QCP in the Lindblad formalism, comparing with the CCP. Section \ref{sect:hpic} examines the Heisenberg picture for the QCP, while Section \ref{sect:Trajectories} examines a QTs approach. Conclusions and outlook are contained in Section \ref{sect:conclusions}.

\section{The Classical and Quantum Contact Processes}
\label{sect:model}

\begin{figure*}[t]
\centering
\includegraphics[width=1\linewidth]{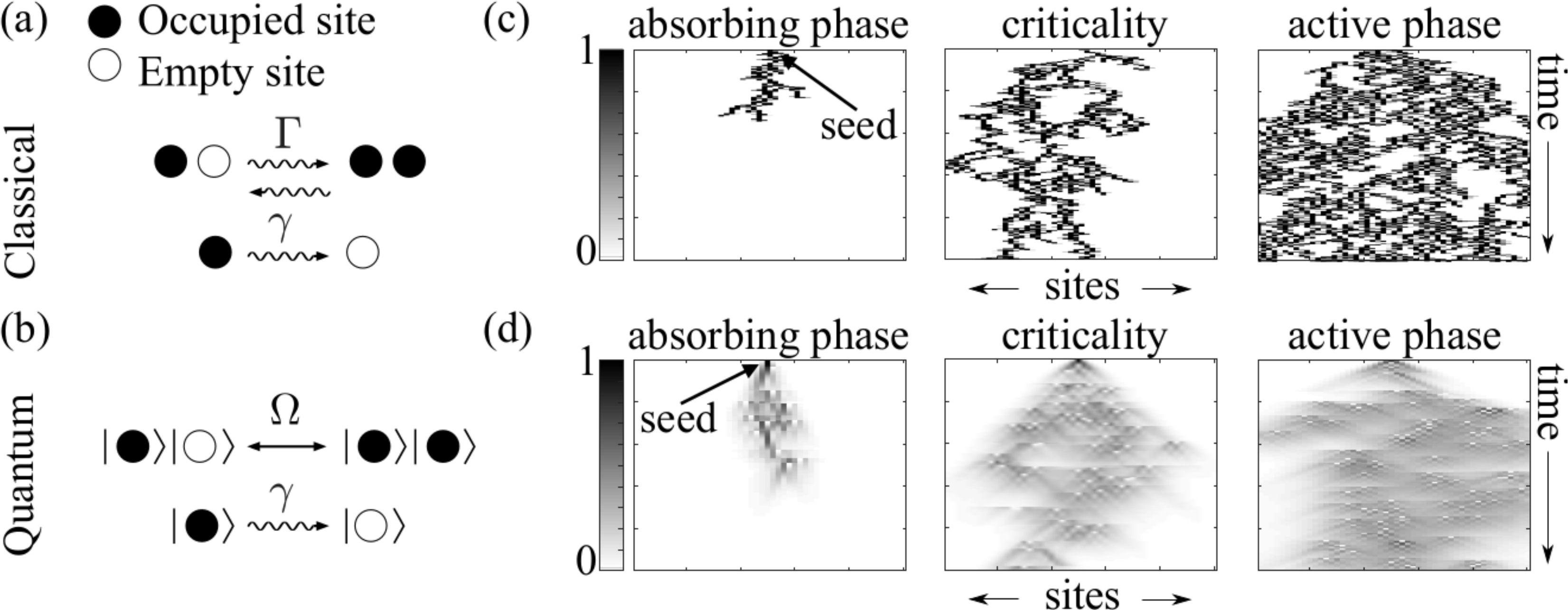}
\caption{\textbf{The Classical and Quantum Contact Processes:}  \textbf{(a)} Dynamical rules for the classical contact process, which consists of branching with strength $\Gamma$ and decay with strength $\gamma$. \textbf{(b)} Corresponding rules for the quantum contact process, consisting of coherent branching with strength $\Omega$ and decay. \textbf{(c)} The classical contact process displays an absorbing state phase transition, as illustrated by the three trajectories shown, one of which is typical of the absorbing phase, one typical of the activate phase and one typical at criticality. Since during a classical trajectory sites can only be empty or occupied with certainty, all squares are either black or white. \textbf{(d)} As with the classical contact process, the quantum contact process displays an absorbing state phase transition. This is illustrated again by three typical trajectories. These show a similar behaviour to the corresponding classical trajectories, as expected by the construction of the dynamical rules for the quantum process. However, the coherent branching in the quantum case leads to expected occupancies different from zero or one.}
\label{fig:contact_process_schematic}
\end{figure*}

\subsection{Lindblad Formalism for Open Quantum Dynamics}

In many physically relevant scenarios, it is not possible to characterise the time-evolution of quantum system by means of unitary closed dynamics. In fact, the simple presence of thermal surroundings or of stochastic processes, e.g. measurements of the system or of its output, makes the study of the time-evolution of the system much more involved. These open quantum systems (OQSs) are not described by usual Schr\"{o}dinger equation. Instead, their average dynamics is implemented, in the Markovian regime, by means of Lindblad-type master equations. These describe the evolution of the average quantum system state $\rho(t)$, where average might mean expectation over a stochastic process or over the degrees of freedom of an environment, according to the physical scenario one has in mind. 

Concretely, the quantum state of the system $\rho(t)$ obeys the Lindblad differential equation \cite{Lindblad1976},
\begin{equation}
\frac{d}{dt}\rho(t)=\mathcal{L}[\rho(t)]\, ,
\label{Lind}
\end{equation}
where the map $\mathcal{L}$, also known as the Lindblad generator, is made of two different contributions,
$$
\mathcal{L}[X]=-i[H,X]+\mathcal{D}[X]\, .
$$
The first piece represents the usual commutator with the system Hamiltonian, which implements the coherent part of the dynamics. On the other hand, the term $\mathcal{D}$, often called the dissipator, contains information about the effects of the environment, or of stochasticity in general, on the system. This must have the following form,
$$
\mathcal{D}[X]=\sum_\mu \left(J_\mu XJ_\mu^\dagger -\frac{1}{2}\left\{X, J_\mu^\dagger J_\mu\right\}\right)\, ,
$$
with $\{A,B\}=AB+BA$, in order to guarantee a physical meaning for the evolved quantum state $\rho(t)$. Note that the dynamics implemented by \eqref{Lind} does not preserve the purity of states and, therefore, in general $\rho(t)$ will represent a mixed state.

\subsection{Model Definitions}

As illustrated in Fig. \ref{fig:contact_process_schematic}, in a contact process, lattice sites can be either occupied ($\ket{\bullet}$) or empty ($\ket{\circ}$) and evolve under processes of spontaneous decay and branching. The CCP, i.e. a contact process with classical branching [Fig. \ref{fig:contact_process_schematic}(a)], can be represented in the Lindblad formalism, thus representing the CCP as an OQS. The evolution of the CCP is then given by a Lindblad master equation with a zero Hamiltonian and dissipative contribution given by two terms,
\begin{align}
\mathcal{D}_{\text{cl}}[\rho] = \mathcal{D}[\rho]+\mathcal{D}_{\rm br}[\rho] ~.
\end{align}
For a system of size $L$, the dissipative term $\mathcal{D}$ is defined as,
\begin{align}
\mathcal{D}[\rho]=\gamma \sum_{k=1}^L\left(\sigma_-^{(k)}\rho\sigma_+^{(k)}-\frac{1}{2}\left\{n^{(k)},\rho\right\}\right)\, ,
\end{align}
where $\sigma_{-}\ket{\bullet}=\ket{\circ}$, $\sigma_{-}\ket{\circ}=0$, $\sigma_{+} = (\sigma_{-})^{\dagger}$, $n$ is the excitation density operator, $n\ket{\bullet}=\ket{\bullet}$, $n\ket{\circ}=0$,  and the parameter $\gamma$ sets the strength of the spontaneous decay. The term $\mathcal{D}_{\rm br}$ instead describes the classical branching/coagulation process and can be written as, 
\begin{equation}
\begin{split}
\mathcal{D}_{\rm br}[X]&=\Gamma\sum_{k=1}^{L-1}J^{(k)}_{\rm L}X(J^{(k)}_{\rm L})^\dagger-\frac{1}{2}\{(J^{(k)}_{\rm L})^\dagger J^{(k)}_{\rm L},X\}+\\
&+\Gamma\sum_{k=1}^{L-1}J^{(k)}_{\rm R}X(J^{(k)}_{\rm R})^\dagger-\frac{1}{2}\{(J^{(k)}_{\rm R})^\dagger J^{(k)}_{\rm R},X\}\, ,
\end{split}
\end{equation}
where $J^{(k)}_{\rm L}=\sigma_1^{(k)}n^{(k+1)}, J^{(k)}_{\rm R}=\sigma_1^{(k+1)}n^{(k)}$ with $\sigma_{1}\ket{\bullet/\circ}=\ket{\circ/\bullet}$.

In the limit $L \to \infty$, the CCP exhibits a NEPT between a unique steady state devoid of particles and a degenerate steady state with finite particle density \cite{Hinrichsen2000}. The former steady state is known as the absorbing state, as once it is reached during the dynamics it cannot be escaped, and the corresponding phase is called the absorbing or inactive phase.

Absorbing phases occur when the strength of the decay process, characterised by the parameter $\gamma$, is sufficient to overcome the process of branching, characterised  by the parameter $\Gamma$. Conversely, when the process of branching is strong enough relative to decay, i.e. the dimensionless branching is large, $\Gamma/\gamma \gg 1$, a steady state with finite particle density can been found, in what is known as the active phase.

In the QCP, branching is coherent and represented by the Hamiltonian,
\begin{align}
H=\Omega \sum_{k=1}^{L-1}\left( \sigma_1^{(k)}n^{(k+1)}+ n^{(k)}\sigma_1^{(k+1)}\right) ~.
\label{eqn:H}
\end{align}
The parameter $\Omega$ thus sets the strength of coherent evolution and, as a consequence, of the branching/coagulation process. The relevant dimensionless parameter is then $\Omega/\gamma$, which determines the relative strength of coherent evolution, tending to create excitations, and dissipation, represented by $\mathcal{D}$, tending to remove them.

As with the CCP, in the limit $L \to \infty$, the QCP displays a non-equilibrium absorbing state phase transition \cite{Marcuzzi2016,Buchhold2017,Roscher2018}. In fact, below the critical points, $\Omega < \Omega_{c}$ and $\Gamma<\Gamma_{c}$, in both models the stationary state is the same unique absorbing state, $\rho_{\text{a}} = \bigotimes_{k} \emptyDM_{k}$. Above the critical point, instead, the stationary state is degenerate and has a finite density of particles. Thus, both models have an active phase, though the stationary state in this phase will differ between the CCP and the QCP.

\subsection{Non-equilibrium Setting}
The specific setting we investigate is that of an initial single seed state evolving under the contact process dynamics \cite{Henkel2008}, see Fig. \ref{fig:contact_process_schematic}. In this scenario, the initial state of the system is a product state. This state is unoccupied at all sites except the central one, $x_{\text{seed}} = \text{floor}(L/2) + 1$. Therefore, the initial state can be written as, $\rho(0) = \sigma_{+}^{(x_{\text{seed}})} \rho_{\text{a}} \sigma_{-}^{(x_{\text{seed}})}$. 

Starting from an initial seed state, the contact process  can be characterised by the survival probability, $P_{\text{sur}}$, total density, $N_{\text{a}}$, and seed-site density, $n_{\text{seed}}$, defined as:
\begin{align}
P_{\text{sur}}(t) &= 1 - \Tr\left[\rho(t)\rho_{a}\right]\, , \\
N_{\text{a}}(t) &= \sum_{k} n(t,k) ~, \\
n_{\text{seed}}(t) &= n(t,k=x_{\text{seed}}) ~,
\label{eqn:obs_defs}
\end{align}
where $n(t,k) = \Tr\left[\rho(t)n^{\left(k\right)}\right]$ is the density profile.

As illustrated in Figs. \ref{fig:contact_process_schematic}(c) and \ref{fig:contact_process_schematic}(d), in the absorbing phase, all clusters generated from a single seed die out so that $P_{\text{sur}}(t), N_{a}(t)$ and $n_{\text{seed}}(t)$ all tend to zero as $t \to \infty$. In contrast, within the active phase in the limit $L \to \infty$, these observables tend to non-zero values as $t \to \infty$. At criticality, the observables are characterised by universal power-law behaviour, which defines the exponents $\delta$, $\Theta$ and $z$ as:
\begin{align}
P_{\text{sur}}(t) &\sim t^{-\delta} , \label{eqn:crit_scaling_1} \\
N_{\text{a}}(t) & \sim t^{\Theta} , \label{eqn:crit_scaling_2} \\
n_{\text{seed}}(t) & \sim t^{\Theta - 1/z} ~.
\label{eqn:crit_scaling_3}
\end{align}

\subsection{Universality Classes of the Contact Processes}

While the CCP has been established to belong to the $1d$ DP universality class \cite{Hinrichsen2000}, the universal properties of the QCP -- in particular the values of exponents and to which class the QCP belongs -- is under debate. However, there have been a number of studies performed from different perspectives. For instance, the QCP has been examined from the perspective of mean-field theory \cite{Marcuzzi2016, Minjae2019}, functional renormalisation group \cite{Buchhold2017, Roscher2018} and tensor networks \cite{Carollo2019}. The transition has been argued to be continuous \cite{Roscher2018}, with the critical point being estimated as $\Omega_{c} \approx 6\gamma$, \cite{Carollo2019}. A selection of critical exponents for the QCP have been estimated in \cite{Carollo2019} as well as in \cite{Buchhold2017}, where the latter includes the effects of classical branching. Table \ref{table:exponents} collects a number of these estimated exponents for the QCP, as well as those of the DP universality classes for comparison. We also emphasise that the QCP can in principle be experimentally realised in Rydberg quantum simulators \cite{Marcuzzi2016,Bloch2012,Bernien2017,Kim2018,Barredo2018}, providing a possible check for theoretical results in the future.

\begin{table}
  \centering
  \begin{tabular}{|c|c|c|c|c|c|}
  \hline
    &QCP & $1d$ DP& $2d$ DP & $1d$ Ref.~\cite{Buchhold2017} & $2d$ Ref.~\cite{Buchhold2017} \\
    \hline
    $\delta$ &$ 0.26 \pm 0.04$&0.16&0.45&$-$&$-$\\
    \hline
    $z$&$ 1.61 \pm 0.16$ &1.58&1.77&1.93&1.97\\
    \hline
    $\Theta$& $0.26 \pm 0.05$ &0.31&0.23&$-$&$-$\\
    \hline
    $\alpha$ &$0.36$ $\pm$ $0.08^{*}$ &0.16&0.45&0.21&0.35\\
    \hline    
  \end{tabular}
  \caption{Relevant exponents of the QCP and CCP. The exponents $\delta, z$ and $\Theta$ can be associated to non-equilibrium observables when starting from an initial seed-state, see \eqref{eqn:crit_scaling_1} - \eqref{eqn:crit_scaling_3}. The exponent $\alpha$ can be associated to the decay of excitation density when starting from a homogenous fully-occupied state, \cite{Henkel2008}. The exponents of the CCP are given by the DP universality class \cite{Hinrichsen2000}. The QCP estimates of $\delta, z$ and $\Theta$ are given in this work (see Section \ref{sect:Trajectories}) with two standard errors, while the value of $\alpha$ indicated by an asterisk is estimated in \cite{Carollo2019}. The two final columns give exponents for the QCP with the inclusion of classical branching at the tricritical point where quantum and classical branching compete, estimated using functional renormalisation group \cite{Buchhold2017}. Note that for DP $\alpha = \delta$, a consequence of the rapidity reversal symmetry that is characteristic of the class \cite{Henkel2008}.} \label{table:exponents}
\end{table}

\section{Time-Evolving Block Decimation in Closed Quantum Systems}
\label{sect:method}


The numerical simulations we perform are based on MPSs and the Time-Evolving Block Decimation (TEBD) algorithm, \cite{Vidal2004}, which we briefly summarise in this section. The TEBD algorithm has been applied extensively in closed quantum systems, on which we focus. 

We consider many-body quantum systems (in $1d$) made of $L$ equal components with single-site Hilbert space having dimension $K$. 
In TEBD, the state $\ket{\psi}$ of such a system is represented as an MPS \cite{Schollwock2011}, 
\begin{align}
\ket{\psi} = \sum_{s_{1},s_{2},...,s_{L}=1}^{K} \mathbf{M}_{s_{1}}^{[1]}\mathbf{M}_{s_{2}}^{[2]}...\mathbf{M}_{s_{L}}^{[L]}\ket{s_{1}s_{2}...s_{L}},
\end{align}
where $\mathbf{M}_{s_{k}}^{[k]}$ are matrices of size $\chi_{k-1} \times \chi_{k}$. The total number of parameters in the MPS representation is $\mathcal{O}\left(L K \chi^{2}\right)$, where $\chi$ is the maximum matrix dimension across the system and is known as the bond dimension. In principle, any state in the many-body Hilbert-space $\mathcal{H}=\mathbb{C}^{K \, \otimes L}$, can be represented exactly as an MPS with $\chi \le K^{\text{floor}(L/2)}$. 

In general, the exact representation of a state as an MPS is not possible, as the number of parameters increases exponentially with $L$. Nonetheless, frequently, relevant quantum states belong to set of states which can be represented as an MPS with $\chi = \mathcal{O}\left(\text{poly}\left(L\right)\right)$, or approximated to some arbitrary accuracy by such an MPS \cite{Eisert2013}. These are said to be represented/approximated efficiently by MPSs. 

As a trivial example, product states can be represented by MPSs with $\chi = 1$. More generally, the $L$ dependence of the von Neumann entanglement entropy,
\begin{align}
S(\varrho_{A}) = - \text{tr}\left[\varrho_{A} \log \varrho_{A}\right] = - \text{tr}\left[\varrho_{B} \log \varrho_{B}\right] ,
\end{align} 
where $\varrho_{A/B}$ is the reduced density matrix of subsystem $A/B$ generated by a bipartition of the system,
$$
{\varrho}_{A/B}=\text{tr}_{B/A} \ket{\psi}\bra{\psi}\, ~ ,
$$
can be used to characterise the efficiency of an MPS representation. In one spatial dimension, an entanglement area law, $ S \sim L^{0}$, for a given state suggests that the state can be efficiently represented as an MPS. This is indeed true for the ground-states of gapped local Hamiltonians, \cite{Hastings2007}, or states with exponentially decaying correlations, \cite{Brandao2015}. However, generally the scaling of $S$ alone is not enough to establish the accuracy of an efficient MPS approximation (in fact all Renyi entropies with index $\alpha < 1$ must be used \cite{Schuch2008,Huang2019}). Nonetheless, the entanglement entropy is useful in practice, particularly as a number of physical systems have been shown to obey area laws \cite{Eisert2010}.

Assuming the initial state of the system to be one represented efficiently by an MPS, then an MPS representation of the time-evolved state can be found by applying some set of operators \cite{Schollwock2011}, typically those constituting a Trotter decomposition of a quantum dynamics \cite{Hatano2005}. When these operators are applied, the resulting MPS will generally have a larger bond dimension than before. In fact, over time this will lead to an exponential increase in the required value of $\chi$ and generally time-evolution can only be treated exactly with MPS for short times \cite{Osborne2006}. This can be linked to the scaling of $S$ with time: If the entanglement entropy is growing linearly with time, as is the case in common quantum quench scenarios \cite{Calabrese2007}, an efficient MPS approximation of the exact state is impossible \cite{Schuch2008}. 

Given the build-up of bond-dimension in an MPS representation over time, the key to performing time-evolution with MPSs is to repeatedly approximate the time-evolved MPSs, thus keeping the total number of parameters under control.

To achieve this, consider the MPS approximation to the state at time $t$, $\ket{\psi_{t}}$, and assume it has a bond dimension $\chi$, which is the maximum we will allow. To approximate the state at time $t'$, we then apply the operator, $\hat{O}$, so that $\ket{\psi_{t'}} = \hat{O}\ket{\psi_{t}}$. The new state $\ket{\psi_{t'}}$ will now have some higher bond-dimension $\chi'$, beyond the maximum we allow in our simulation. To remedy this, we want to find an MPS approximation to $\ket{\psi_{t'}}$ that has bond dimension $\chi$. Calling this approximation $\ket{\phi_{t'}}$, we then want to solve the minimisation problem,
\begin{align}
\ket{\phi_{t'}} = \text{argmin}_{\ket{\phi}}  |\ket{\psi_{t'}} - \ket{\phi}| ,
\label{eqn:truncation_problem}
\end{align}
where $| \cdot |$ indicates the Hilbert-space norm. One can iterate this procedure to produce an approximation to the time-evolution of the state, allowing one to calculate desired observables along the way \cite{Schollwock2011}.

The TEBD algorithm is, in essence, a simple approximation to the solution of $\left(\ref{eqn:truncation_problem}\right)$ given by considering successive bipartitions of the system at $k = 1, 2, ..., L$. At each cut, one performs a Schmidt decomposition of the state $\ket{\psi_{t'}}$ and discards a sufficient number of the smallest Schmidt coefficients so as to reduce the bond-dimension to $\chi$. More specifically, across the cut at site-$k$ the Schmidt coefficients, $\lambda_{j}^{[k]}$, are placed into a non-ascending order $\lambda_{1}^{[k]} \ge \lambda_{2}^{[k]} \ge ... > 0$. The MPS approximation with bond-dimension $\chi$ at site $k$ is then constructed by retaining the $\chi$ largest Schmidt coefficients. In other words, all Schmidt coefficients with $j \ge \chi + 1$ are discarded, and the corresponding discarded weight,
\begin{align}
\epsilon_{\chi}^{[k]} = \sum_{j \ge \chi+1} \left(\lambda_{j}^{[k]}\right)^{2} ,
\label{eqn:discarded_weight}
\end{align}
measures the error in this approximation. At any given cut, this approximation is optimal and so overall this provides a simple approximation to the solution of (\ref{eqn:truncation_problem}). We will use this method throughout, though we remark that many more sophisticated variations exist \cite{Paeckel2019}.

\section{Simulation of Universal Dynamics in the Double-Space}
\label{sect:simulation_double_space}

\subsection{Double-space Representation of Lindblad Dynamics}

Perhaps the most straightforward way to apply TEBD to the study of Lindblad dynamics, (\ref{Lind}), is to represent the density matrix $\rho(t)$ as a vector, $\ket{\rho(t)}$, in the ``double-space" defined via the Choi-isomorphism, $\ket{n}\hspace{-3pt}\bra{m} \to \ket{n}\otimes\ket{m}$, \cite{Choi1975}. One thus has, 
$$
\rho(t)=\sum_{mn}\rho_{mn}(t)\ket{m}\bra{n}\to \ket{\rho(t)}=\sum_{mn}\rho_{mn}(t)\ket{m}\otimes \ket{n}\, .
$$

Mapped in this way, the evolution of the quantum state can then be shown to be generated by the following Schr\"{o}dinger-like equation ,
\begin{align}
\frac{d}{dt}\ket{\rho(t)}=\mathbb{L}\ket{\rho(t)}\, ,
\label{eqn:double_space_Lindblad}
\end{align}where $\mathbb{L}$ is the representation of the Lindblad map \eqref{Lind} in the double space. It is possible to show that one must have, 
\begin{align}
\mathbb{L} = -i H_{D} + \mathcal{D}_{D}  ~ , 
\end{align}
where $H_{D}$ is Hermitian and has the form,
\begin{align}
H_{D} = \left(H \otimes \mathds{1} - \mathds{1} \otimes H^T \right) ~ ,
\end{align}
while,
\begin{align}
\mathcal{D}_{D} = \sum_\mu \left(J_\mu\otimes J_\mu^* -\frac{1}{2}J_\mu^\dagger J_\mu\otimes \mathds{1} -\frac{1}{2}\mathds{1}\otimes J_\mu^T J_\mu^*\right)\, ,
\end{align}
where $^*$ means complex conjugation and $^T$ matrix transposition. 

Solutions to (\ref{eqn:double_space_Lindblad}) give the evolved state up to time $t$, $\ket{\rho(t)}=e^{t\mathbb{L}}\ket{\rho(0)}$, where $\ket{\rho(0)}$ is the initial condition for the density matrix in the vectorized representation. By performing a Trotter decomposition of the time-evolution operator $e^{t \mathbb{L}}$, which we choose to be a second-order scheme, one can apply the TEBD algorithm naturally to approximate $\ket{\rho(t)}$ using an MPS with bond-dimension $\chi$. From the approximation of $\ket{\rho(t)}$, observables can then be calculated as
\begin{align}
O(t) &= \text{tr}\left[ \rho(t) \hat{O}\right] = \braket{\mathds{1} | \hat{O}_{D} | \rho(t)} ~ ,
\end{align}
where $\hat{O}_{D} = \hat{O} \otimes \mathds{1}$ and $\ket{\mathds{1}}$ is the double-space state representation of the identity operator. 

When considering the MPS representation of $\ket{\rho(t)}$, a natural quantity to consider is the operator space entanglement,
\begin{align}
\tilde{S} = - \text{tr}\left[\tilde{\varrho}_{A} \log \tilde{\varrho}_{A}\right] = - \text{tr}\left[\tilde{\varrho}_{A} \log \tilde{\varrho}_{A}\right] ,
\end{align} 
with 
$$
\tilde{\varrho}_{A/B}=\text{tr}_{B/A} \ket{\rho}\bra{\rho} ~ .
$$
The value of $\tilde{S}$ plays an analogous role to the von Neumann entropy, $S$, for closed quantum systems and provides a characterisation of the computational difficulty of TN simulations. 

\subsection{Schr\"{o}dinger Picture Results}

\begin{figure*}[t]
\centering
\includegraphics[width=1\linewidth]{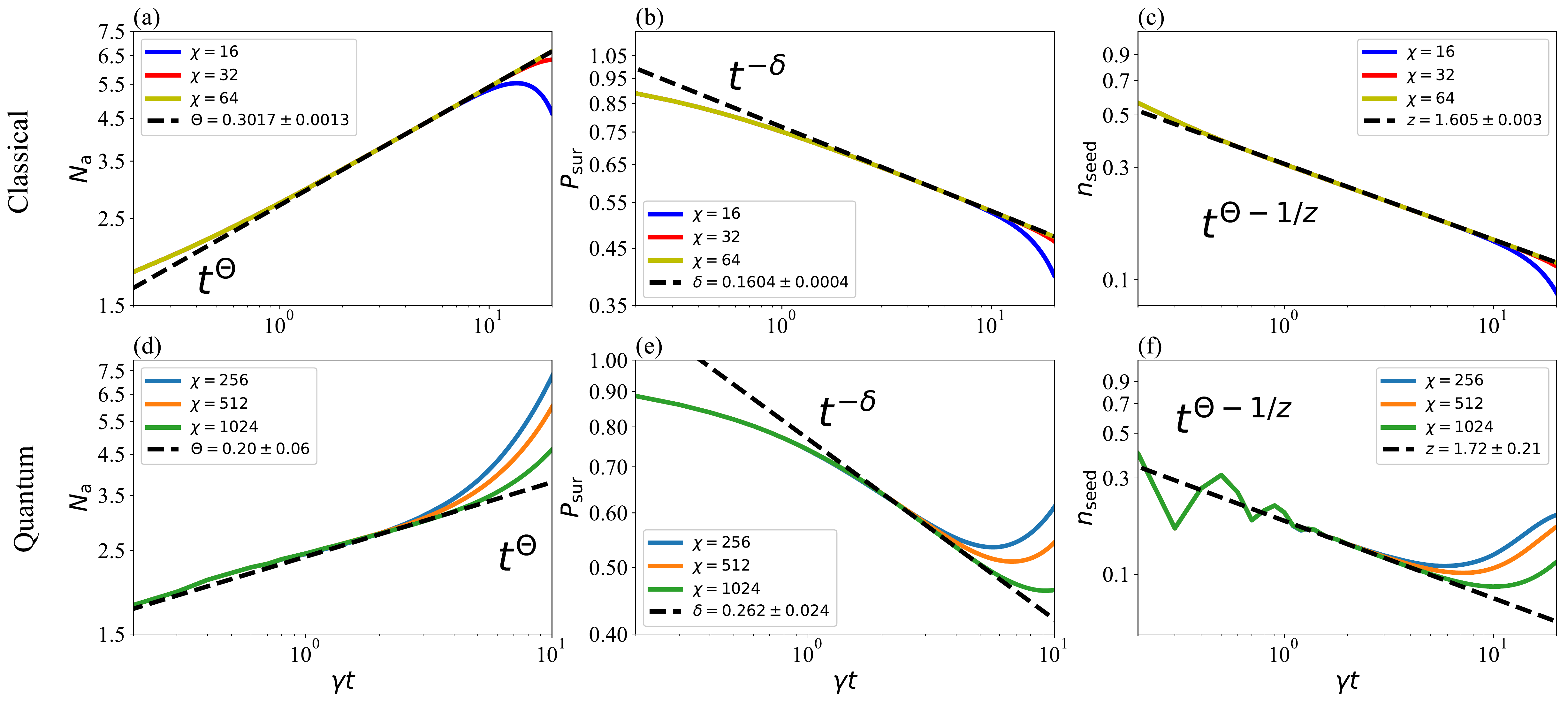}
\caption{\textbf{Critical Dynamics in the Classical Contact Process (a) - (c) :} The plots show the approximations of time-dependent observables for $\Gamma = 6.75 \gamma$, obtained with the TEBD algorithm. \textbf{(a)} The plot shows the total density, $N_{\text{a}}(t)$, calculated using the TEBD algorithm in the double space, with bond-dimensions $\chi = 16,32$ and $64$. At late times, finite bond-dimension effects can be seen clearly. To establish the critical exponents, power-law fits were performed in the interval $\gamma t \in [5,10]$ for $\chi = 32$ and $64$, with the latter fit determining the best estimate of $\Theta$ and the difference between the two used as the error. \textbf{(b)} The plot shows the survival probability, $P_{\text{sur}}(t)$, calculated using the same MPS approximations as for $N_{\text{a}}(t)$. As in that case, significant finite-bond effects can be seen, but again only a modest $\chi = 64$ is needed to accurately approximate the observable. \textbf{(c)} The plot shows the evolution of the seed-site density, $n_{\text{seed}}(t)$, and a power-law fit determines the exponents $\Theta - 1/z$. Combined with the estimated $\Theta$ from (a), this leads to an estimate of $z$, albeit with a relatively high error estimate due to error propagation. \textbf{Critical Dynamics in the Quantum Contact Process (d) - (f):} The plots show approximations of universal observables for the quantum contact process at $\Omega = 6 \gamma$, estimated with TEBD in the double-space. These can be compared with the corresponding plots for the classical contact process, which are estimated using an identical algorithm and show similar qualitative behaviour. See also Fig. \ref{fig:QCP_exponents_trajs} for the same quantities and analysis using a trajectories approach.  \textbf{(d)} The plot shows the total density, $N_{\text{a}}(t)$, with bond-dimensions $\chi = 256, 512$ and $1024$. Compared with the corresponding classical plot, the finite-bond effects are larger and much higher bond-dimensions are required to achieve convergence. \textbf{(e)} The survival probability, $P_{\text{sur}}(t)$. Once again finite $\chi$ effects are significant and to reach $\gamma t = 10$ a bond-dimension of at least $\chi = 2046$ is likely required. \textbf{(f)} The seed-site density evolution $n_{\text{seed}}(t)$ for the QCP. The estimate of the exponent $z$ is obtained, as in the classical case, from a power-law fit within $\gamma t \in [2,4]$, with error propagation leading to a relatively large error estimate for this value.
}
\label{fig:Critical_Exponents_CCP_QCP}
\end{figure*}

The evolution of the total density, survival probability and seed-site density are shown for the CCP and QCP in Fig. \ref{fig:Critical_Exponents_CCP_QCP}. In both cases the same TEBD algorithm is used, with a fixed Trotter step of $\gamma \delta t = 0.1$ and bond-dimensions of $\chi = 16, 32, 64$ and $256, 512, 1024$ for the CCP and QCP respectively. The simulations are performed with $\gamma = 1$ at the estimated critical points, $\Gamma = 6.75 \gamma$ and $\Omega = 6 \gamma$ for the CCP and QCP respectively. For the CCP the critical point was estimated by scanning various values of $\Gamma$ and finding where both the total density and the survival probability show little deviation from a straight line in a log-log plot. In the case of the QCP, we take the previously estimated critical value of $\Omega = 6 \gamma$,  \cite{Carollo2019}. Both sets of observables show the correct qualitative behaviour, in line with the expectations of the critical dynamics, \eqref{eqn:crit_scaling_1} - \eqref{eqn:crit_scaling_3}.

To approximate the critical exponents, power-law fits were performed to $N_{\text{a}}(t)$, $P_{\text{sur}}(t)$ and $n_{\text{seed}}(t)$ thus estimating $\Theta , \delta$ and $z$. To provide a best-estimate of these values, the simulations with largest $\chi$ were used for the fits, while the absolute differences between these estimates and those obtained by fitting to a $\chi$ of half the maximum were used for the error estimates. 

In both the CCP and QCP, while all the different bond-dimension simulations agree at early times, at later times the low bond-dimension runs deviate considerably. This suggests that finite-bond effects can lead to a significant build-up of errors in observables, as in the closed quantum system case, even for classical states. However, the bond-dimensions needed to reach convergence until $\gamma t=20$ are very modest for the CCP. Since the fits for the CCP performed over $\gamma t \in [5,10]$ lead to estimated exponents within a few percent of the true $1d$ DP values, we see that critical dynamics of the CCP can be accurately simulated with MPSs in the double-space, and critical exponents estimated with small errors. 

Compared with the CCP, simulation of the QCP requires much larger values of $\chi$ to achieve convergence in the examined observables, with $\chi = 1024$ showing large deviations from a power-law by $\gamma t = 10$. As such, fitting from $\gamma t \in [2,4]$ was used to establish these exponents, where the $\chi = 1024$ simulations closely follow a power-law. As this is a relatively early time for which to perform the fits, these exponents may well contain finite-time errors. However, for comparison, fitting the CCP from $\gamma t \in [1,5]$ as opposed to $\gamma t \in [5,10]$ only increases the errors in the estimated exponents from around $1\%$ to $5\%$. 

\subsection{Operator Space Entanglement in the Schr\"{o}dinger Picture}
\label{sect:entropy}

\begin{figure}[t]
\centering
\includegraphics[width=1\linewidth]{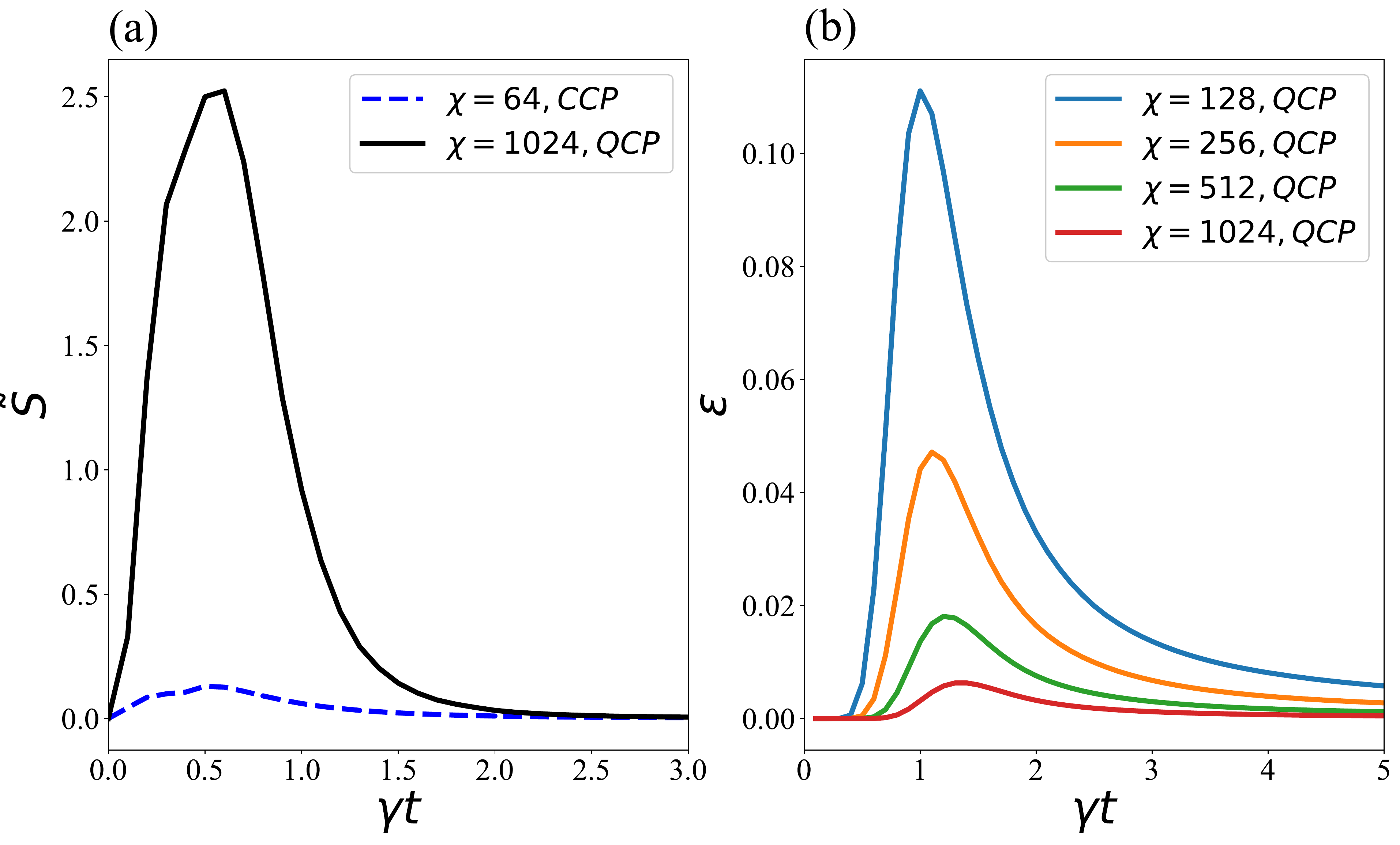}
\caption{\textbf{Evolutions of the Operator Space Entanglement and Discarded Weight:} \textbf{(a)} Operator space entanglement, $\tilde{S}$ for the CCP and QCP. In both cases a barrier like structure can be seen, consistent with the initial build-up of correlations followed by a decrease due to dissipation. The fact that the barrier for the CCP is much lower than for the QCP suggests that the dynamics should be much easier to approximate in the former case. \textbf{(b)} The evolution of the error estimate for the QCP, defined here as the square root of the sum of squared Schmidt coefficients discarded during truncation. As for $\tilde{S}$, the error shows a barrier like structure, consistent with a build-up of errors that lead to the significant finite $\chi$ effects seen in Fig. \ref{fig:Critical_Exponents_CCP_QCP}. As the value of $\chi$ is doubled, the barrier peak decreases correspondingly, leading to more accurate approximation of observables over longer times.}
\label{fig:EntropyOpSpace}
\end{figure}

To understand why the QCP is much harder to simulate than the CCP, we can compare the evolution of the operator space entanglement entropy, $\tilde{S}$. Taking the maximum value of the entanglement entropy across all bipartitions throughout, the evolutions of $\tilde{S}$ for the CCP and QCP are shown in Fig.~\ref{fig:EntropyOpSpace}(a).

In both the CCP and QCP, $\tilde{S}$ shows a clear ``barrier" behaviour, where initially it grows rapidly to a peak around $\gamma t = 0.5$ before decaying to a lower final value. This is consistent with an initial period of branching/coagulation evolution, where correlations build up rapidly and spontaneous decay is irrelevant, followed by a period where the latter becomes relevant and removes correlations/excitations from the state. While the overall picture seems the same for both the CCP and QCP, in the classical case the barrier is clearly much lower than in the QCP. Given that the operator space entanglement entropy should characterise the error in simulations, the difference between $\tilde{S}$ in the CCP and QCP helps explain the difference in accuracies found in observables. The relationship between $\tilde{S}$ and errors is illustrated in Fig.~\ref{fig:EntropyOpSpace}(b), which shows the error in the simulations of the QCP over time, equal to the square root of the discarded weight defined in (\ref{eqn:discarded_weight}). As the value of $\chi$ is increased, the error at each time drops, approximately halving when the bond-dimension is doubled. As with the entropy, the error shows a peak-like structure, with the peak occurring shortly after that of the entropy. 

\section{Heisenberg Picture in the Double-space}
\label{sect:hpic}

To study the Heisenberg picture dynamics with TEBD in the double-space, one takes a representation of an operator $O$ as a vector $\ket{O}$, then evolves it through the dual Lindbladian $\mathbb{L}^\dagger$, such that
$$
\ket{O(t)}=e^{t\mathbb{L}^\dagger}\ket{O}\, .
$$

In the case of closed quantum systems, the Heisenberg picture can be used to extend the maximum time over which simulations are accurate by roughly a factor of two \cite{Paeckel2019}. Intuitively, this is expected because the dual dynamics of the unitary evolution, $U^{\dagger}(t)$, is equal to the original dynamics but backwards in time, $U^{\dagger}(t) = U(-t)$. Thus, one might expect that the dual dynamics is not radically different in terms of computational difficulty than that of the usual dynamics, and both maybe accurately approximated up to the same time, thus doubling the maximum time when combined. 

In contrast, for open quantum systems, the dynamics implemented by $\mathbb{L}^\dagger$ is in principle completely different from the one of $\mathbb{L}$. This is exemplified by the case of the QCP, where the dual dynamics does not have an absorbing state. It has previously been suggested that the absorbing state, which is the unique steady state for any finite-size system, makes the application of tensor network methods more challenging \cite{Carollo2019}. Therefore, one might expect that the absence of the absorbing state in the dual dynamics will mean that applying the Heisenberg picture is more accurate than the Schr\"{o}dinger picture for the QCP. 

\begin{figure}[t]
\centering
\includegraphics[width=1\linewidth]{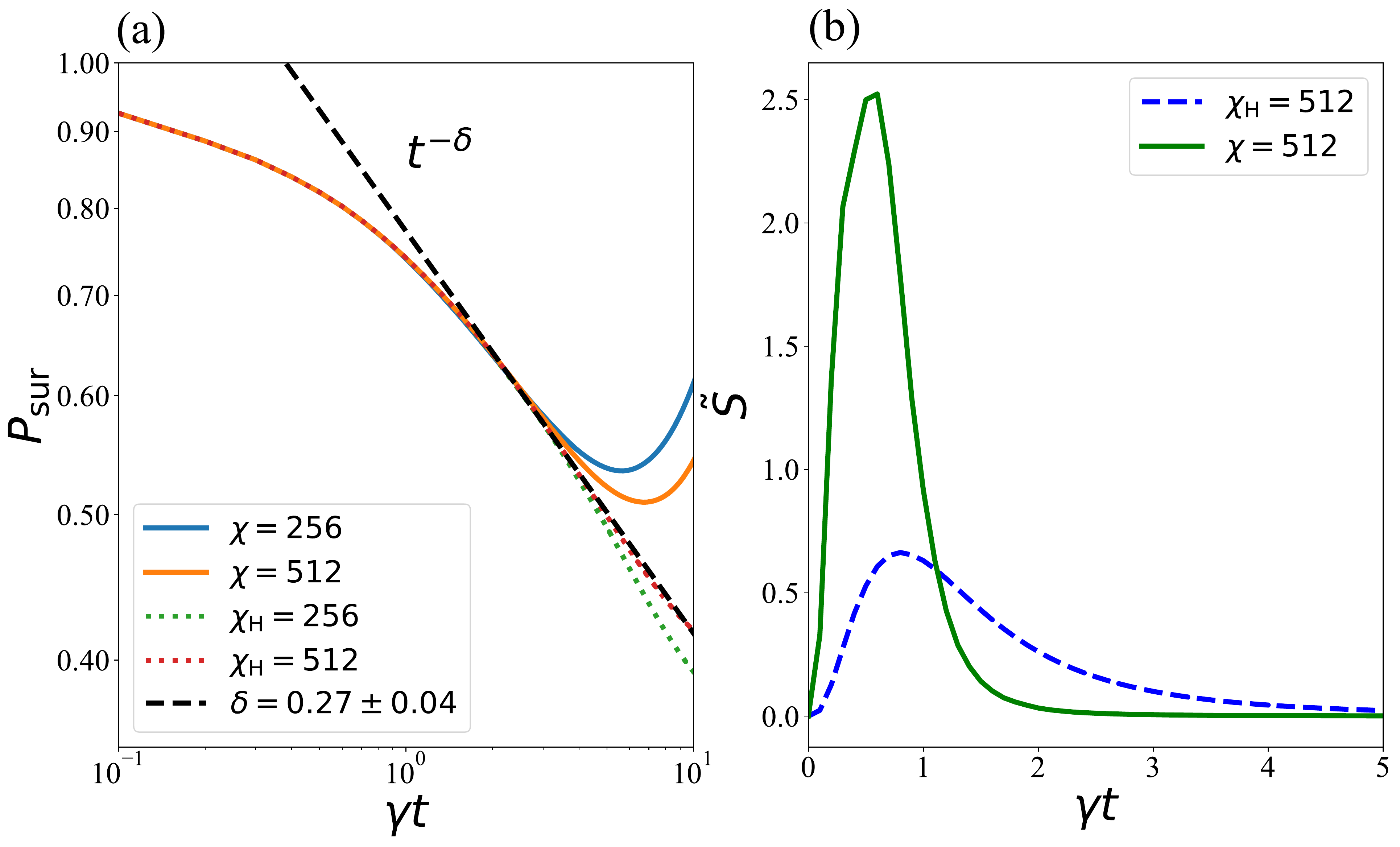}
\caption{\textbf{The Survival Probability and Entanglement in the Heisenberg Picture:} \textbf{(a)} The survival probability, $P_{\text{sur}}$, calculated in the Heisenberg picture, compared with the Schr\"{o}dinger picture for $\chi = 256$ and $512$ (the Heisenberg picture runs are denoted by a subscript H). While the Schr\"{o}dinger picture approximations deviate dramatically from the expected power-law behaviour, the Heisenberg picture simulations seem much better and the $\chi_{\text{H}} = 512$ case can be approximated by power-law until $\gamma t = 10$. \textbf{(b)} The entropy barrier for the dynamics in the Heisenberg picture and Sch\"{o}dinger picture dynamics. In the Heisenberg picture, the evolution of $\tilde{S}$ shows the same barrier structure as for the Schr\"{o}dinger picture. However, the shape of the barrier is different, with a considerably lower peak. This suggests that the same bond-dimension, and therefore the same computational costs, would lead to a significantly higher accuracy approximation, consistent with the results for $P_{\text{sur}}$
.}
\label{fig:Hpic}
\end{figure}

To explore this, we calculate the survival probability for the QCP in the Heisenberg picture using TEBD and $\chi = 256, 512$, along with the operator-space entanglement entropy, as shown in Fig. \ref{fig:Hpic}. As can be seen, the entropy displays a characteristic barrier as in the Schr\"{o}dinger picture. However, the barrier is substantially lower for the Heisenberg picture than for the Schr\"{o}dinger picture case (though it is also less sharply peaked). Correspondingly, the survival probability shows dramatically reduced finite bond-dimension effects, with the $\chi = 512$ approximation showing reasonable power-law behaviour up until $\gamma t = 10$, leading to an estimated exponent of $\delta = 0.27 \pm 0.04$ when fit over $\gamma t \in [2,4]$.  

On a practical level, these results suggest that the Heisenberg picture might allow for a more accurate approximation of the survival probability, $P_{\text{sur}}$, and thus $\delta$, at cheaper computational cost (lower bond-dimension). Other observables such as the total density can also be approximated in the Heisenberg picture, e.g., in order to establish the exponent $\Theta$ with greater accuracy, though we do not discuss this direction further.

\section{Trajectories for the QCP}
\label{sect:Trajectories}

\subsection{Entanglement Distribution}

As an alternative to the Lindblad formalism and double-space approach, we now consider a stochastic unravelling of the master equation realised algorithmically by Quantum Jump Monte Carlo (QJMC) \cite{Plenio1998,Gardiner2004}. In this QTs approach, each individual quantum trajectory corresponds to a pure state evolution. Therefore, the standard TEBD method for closed quantum systems can be applied quite directly to simulate a given trajectory, with the sample means over all trajectories providing an approximation for the observables of the average (Lindblad) dynamics.

As before, we will be interested in quantifying the presence of entanglement in the dynamics and how this affects the accuracy of TN calculations of universal non-equilibrium physics. Since each individual trajectory in QJMC is governed by a pure state evolution, the von Neumann entanglement entropy of the state has the usual physical meaning. However, unlike the  closed system case, the entanglement of a trajectory, $S_{\text{traj}}$, can now be considered a random variable and thus associated to a distribution/probability. While this might seem to be a complicating feature compared to the single operator space entanglement entropy in the double-space case, it is actually very helpful for building a picture of the accuracy of simulations using MPSs: If we associate to a given bond-dimension, $\chi$, some characteristic maximum entanglement ``cutoff", $\bar{S}(\chi)$, then we expect that the set of trajectories with $S_{\text{traj}} \ll \bar{S}(\chi)$ will be well approximated, while those with $S_{\text{traj}} \approx \bar{S}(\chi)$ will be subject to finite bond-dimension effects. In other words, for low-entanglement trajectories the entanglement cutoff will be irrelevant but for high-entanglement trajectories it will be relevant.  

\begin{figure*}[t]
\centering
\includegraphics[width=1\linewidth]{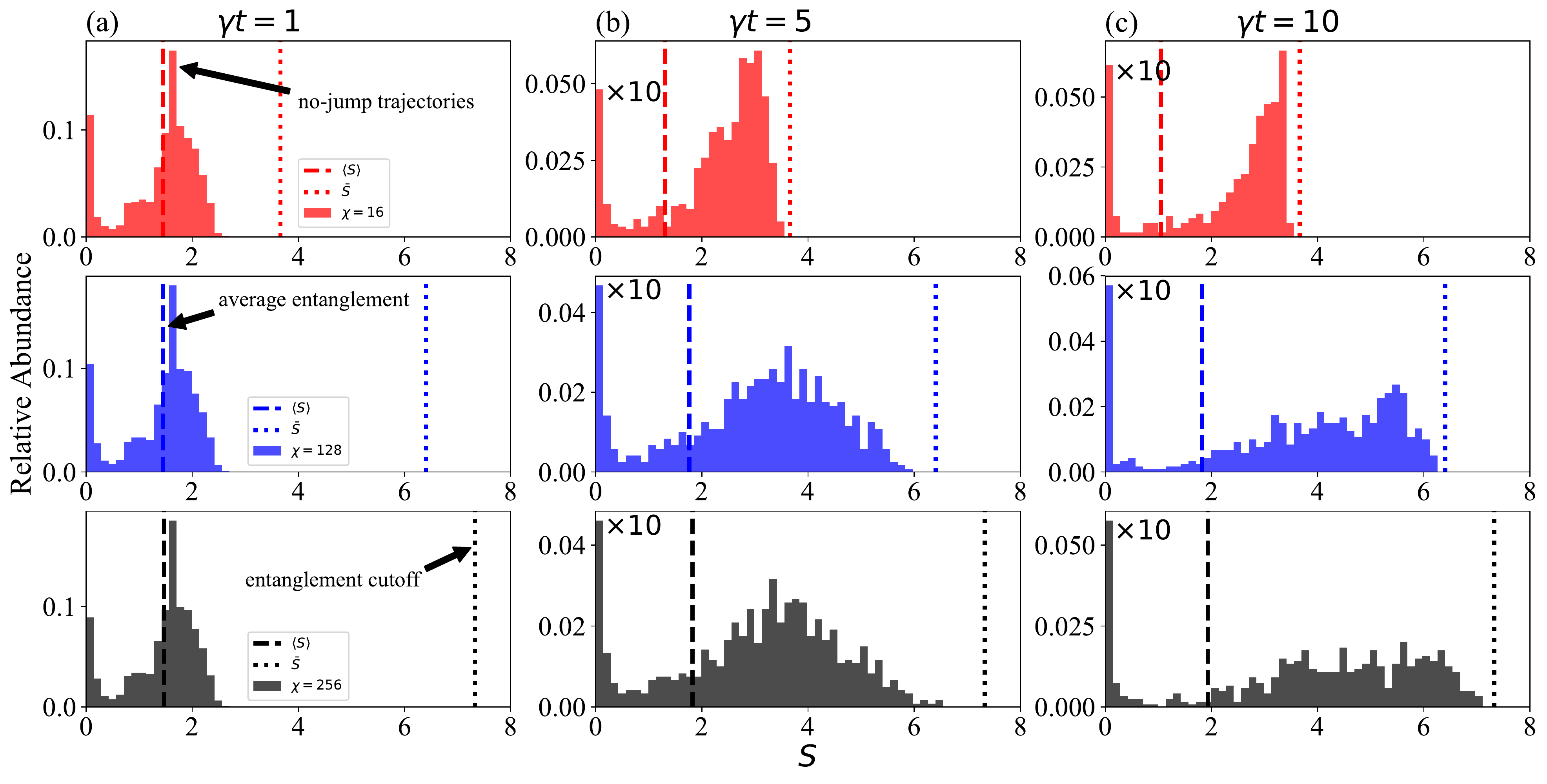}
\caption{\textbf{Empirical Distribution of Entanglement for QJMC Trajectories:} Histograms of $1000$ trajectories generated by QJMC and TEBD with bond-dimensions $\chi = 16,128$ and $256$. The bins are determined by dividing the $\chi = 256$ entanglements at $\gamma t = 10$ into $50$ equal width bins. The histograms are normalised by the total number of trajectories, such that the sum of heights is one. For $\gamma t = 5$ and $\gamma t = 10$, the first bin, which covers the absorbing state with $S = 0$, has been scaled by a factor of $0.1$ to allow clearer visualisation of the distributions at larger values of $S$. The mean values of $S$ as each time, $\langle S \rangle$,  are indicated by vertical dashed lines. The entanglement cutoffs, $\bar{S}$, defined as the maximum values of $S$ for each bond-dimension taken over all times and trajectories,  are indicated by dotted vertical lines. \textbf{(a)} The first column compares the three distributions at $\gamma t = 1$. All distributions agree closely and display a clear bimodal behaviour. The mode near $S=0$ can be interpreted as the trajectories that have fallen into the absorbing state. The second mode, which has a large density around a single value, corresponds to the proportion of trajectories that have not yet jumped and thus have evolved deterministically. \textbf{(b)} The second column displays the three distributions at $\gamma t = 5$. By this time the $\chi = 16$ distribution differs significantly from the others, showing a wall-like behaviour near the corresponding maximum entanglement, while the other distributions largely agree. \textbf{(c)} By $\gamma t = 10$, shown in the third column, the distributions for $\chi = 128$ and $256$ differ visibly. However, the difference represents only a small fraction of weight overall, and one can expect that observables calculated with both $\chi = 128$ and $\chi = 256$ will be similar.}
\label{fig:traj_ent_hist}
\end{figure*}

To explore these issues, we first consider the distribution of $S_{\text{traj}}$ in $1000$ trajectories simulated with TEBD for $\chi = 16,128$ and $\chi=256$. The evolution of each trajectory was calculated up until $\gamma t=10$ using a Trotter-step of $\gamma \delta t = 0.01$, chosen so that the order of the associated error in the QJMC (which is a first-order scheme) is comparable with the second-order scheme used in the double-space case. The entanglement distributions for $\gamma t = 1, 5$ and $10$ are shown in Fig.~\ref{fig:traj_ent_hist}, with the rows illustrating the evolution of the entanglement distributions for  $\chi = 16,128$ and $\chi=256$ respectively. Details of the histogram construction are given in the caption.

\begin{figure*}[t]
\centering
\includegraphics[width=1\linewidth]{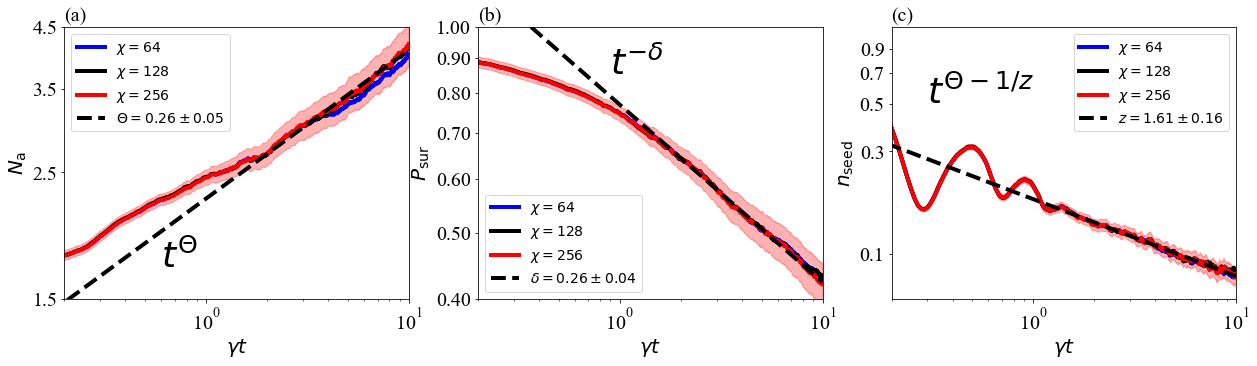}
\caption{\textbf{Critical Dynamics in the QCP using the Trajectories Approach:} Plots showing the evolution of $N_{\text{a}}(t), P_{\text{sur}}(t)$ and $n_{\text{seed}}(t)$, calculated using QJMC and TEBD (see the second row of Fig. \ref{fig:Critical_Exponents_CCP_QCP} for the same quantities calculated using the double-space approach). The error estimates for the exponents are calculated by bootstrap: The relevant power-laws are refit to $1000$ datasets of $1000$ trajectories generated by resampling. The error is then given as twice the standard deviation of the resulting empirical distributions. \textbf{(a)} The evolution of $N_{\text{a}}(t)$ calculated from the sample mean of $1000$ trajectories with $\chi = 64, 128$ and $256$. The shaded region indicates the statistical uncertainty for the $\chi = 256$ estimate, quantified as twice the standard error. All three curves lie within this region, indicating that finite bond-dimension effects are small relative to statistical error. All curves show a roughly power-law behaviour up to $\gamma t = 10$, a significant improvement over the double-space case (Fig. \ref{fig:Critical_Exponents_CCP_QCP}). The critical exponent $\Theta$ was estimated by fitting a power-law between $\gamma t \in [5,10]$ (shown as the dashed black line). \textbf{(b)} The evolution of $P_{\text{sur}}(t)$, calculated using the same method as $N_{\text{a}}(t)$. Once again all three curves lie well within the shaded region and display an approximate power-law behaviour and fitting between $\gamma t \in [5,10]$ establishes $\delta$. \textbf{(c)} The evolution of $n_{\text{seed}}(t)$. As with $N_{\text{a}}(t)$ and $P_{\text{sur}}(t)$, the convergence with bond-dimension and agreement with a power-law is dramatically improved relative to the double-space results of Fig. \ref{fig:Critical_Exponents_CCP_QCP}. However, the presence of statistical errors still leads to a relatively high uncertainty on the estimate of the critical exponent $z$, calculated from the power-law fit to $\Theta - 1/z$ and using the value of $\Theta$ from (a).} 
\label{fig:QCP_exponents_trajs}
\end{figure*}

In Fig.~\ref{fig:traj_ent_hist}(a), when $\gamma t=1$, all the histograms displayed are very similar; they show the same bimodality with one peak at $S=0$ - attributable to the unentangled absorbing state - and are similar except for the fact that the lower bond-dimension distributions display slightly more weight near $S=0$. As such, the means of the distributions, $\langle S \rangle$, shown as vertical dashed lines, are very close. This similarity can be explained by comparing these distributions with the maximum value of $S$ found for any trajectory at any time. These values, plotted as dotted vertical lines, can be considered as a measurement of the cutoff, $\bar{S}$. For all $\chi = 16, 128$ and $256$, the values of $\bar{S}$ lie well above the support of the distribution, and we can expect the effect of the cutoff to be minimal. In fact, given the separation between the support and $\bar{S}$, we might expect there to be essentially no effect and it is interesting that there is still a clear discrepancy at $S = 0$. This discrepancy again suggests that the presence of an absorbing state poses challenges for numerical simulations.

In Fig.~\ref{fig:traj_ent_hist}(b), where $\gamma t=5$, the distribution of entropies for $\chi=16$ and $\chi=256$ differ significantly, with a large proportion of trajectories for $\chi = 256$ displaying entropies larger than the entanglement cutoff for $\chi=16$, leading to an artificial build-up of weight around $\bar{S}$ for the $\chi = 16$ simulations. Thus, we can conclude that the finite-entanglement cutoff is relevant at this time for $\chi = 16$, and the effect on the mean is clearly visible. This is in contrast to the case of $\chi = 128$, which retains a good agreement with the $\chi = 256$ simulations.

In Fig.~\ref{fig:traj_ent_hist}(c), where $\gamma t=10$, the support of the distribution for $\chi = 256$ is close to the entanglement cutoff $\bar{S}(\chi = 256)$, and above $\bar{S}(\chi = 128)$. This corresponds to a noticeable difference in the entanglement distributions for $\chi = 128$ and $\chi =256$, though the difference is only slight compared to that of $\chi = 16$. In fact, since only a small weight appears in the $\chi = 256$ distribution above the value of $\bar{S}(\chi = 128)$, we might expect that the accuracy of the $\chi = 128$ simulations for observables will still be reasonably good. Moreover, we might expect that the $\chi = 256$ simulations themselves are accurate due to the fact there is no significant build-up of weight near $\bar{S}(\chi = 256)$, which for the $\chi = 16$ case provided a clear indication of the entanglement cutoff's relevance.

\subsection{Universal Dynamics with Trajectories}

To assess the accuracy of the trajectories approach for the QCP and investigate the relationship between the entanglement distributions and the approximation of observables, we repeat the analysis for the QCP performed in Section \ref{sect:simulation_double_space}, as shown in Fig.~\ref{fig:QCP_exponents_trajs} for $\chi =  64, 128$ and $256$. From analysis of the entanglement distributions, Fig. \ref{fig:traj_ent_hist}, we expect to find good accuracy up to $\gamma t=10$ for $\chi = 128$ and $256$, and indeed the curves of $N_{\text{a}}(t),  P_{\text{sur}}(t)$ and $n_{\text{seed}}(t)$ overlap closely for these bond-dimensions, with $\chi = 64$ deviating more noticeably. 

Compared to the double-space simulations, Fig.~\ref{fig:Critical_Exponents_CCP_QCP}, the observables calculated from trajectories seem much more accurate; with $\chi = 128$ and $256$ they converge up to $ \gamma t = 10$ within statistical errors (given as two standard errors from the sample-mean and indicated by the shaded region in plots). Furthermore, they display the expected power-law behaviours, \eqref{eqn:crit_scaling_1} -- \eqref{eqn:crit_scaling_3}, for this whole period. This allows for fits to be performed over a longer region of time, $\gamma t \in [2,10]$, thus helping to eliminate finite-time errors. Since the curves for $\chi = 64, 128$ and $\chi = 256$ lie well within the same shaded region for each observable, we can conclude that the finite bond-dimension effects are essentially negligible relative to the statistical errors in this region. As such, we provide purely statistical error estimates on the estimated exponents, obtained via a statistical bootstrap (see Fig. \ref{fig:QCP_exponents_trajs} for details). While these error estimates are large (corresponding to an approximate $95\%$ confidence interval), they can easily be reduced by increasing the number of samples. 

\section{Conclusions and Outlook}
\label{sect:conclusions}

In this paper, we have made use of MPSs to study the critical dynamics of the classical and the quantum contact processes, which display non-equilibrium absorbing state phase transitions. For the QCP we have shown how the Heisenberg picture, where the dynamics does not display an absorbing state, can be used to improve the accuracy of simulations in the Lindblad formalism, over and above the Schr\"{o}dinger picture. Moreover, when combining TEBD with a Quantum Jump Monte Carlo approach, we find that the expected critical behaviour can be reproduced with much higher accuracy for longer times than that of the Schr\"{o}dinger picture Lindblad approach. In all approaches, we show that the entanglement in the MPSs can be used to understand these differences clearly, providing both a useful diagnostic tool and physical picture that links the numerical method to the dynamics in question.

The difference in accuracies found between the Lindblad and QTs approaches in the case of the critical QCP emphasises, as has been mentioned previously \cite{Jaschke2018}, that when considering the simulation of open quantum systems with TN, one should examine different approaches to simulating the dynamics carefully. Given the observed superiority of a QTs approach for capturing the critical QCP dynamics, it would be interesting to know if this result is more general and whether approaches such as QJMC can allow one to study critical dynamics in other systems at higher accuracies than possible otherwise.  

Finally, in the process of investigating these primary issues, we have also provided several results for the critical physics of the QCP. The most convincing conclusion that we can draw from these is that the universality class of the QCP cannot be DP, as evidenced by the fact that the best estimate of the exponent $\delta = 0.26$ is far from that of $1d$ DP and $2d$ DP, see Fig. \ref{fig:QCP_exponents_trajs} and Table \ref{table:exponents}. This was also confirmed by the results from the double-space calculations, shown in Fig. \ref{fig:Critical_Exponents_CCP_QCP}. However, since that finite bond-dimension errors are small relative to statistical errors when using QJMC, we can use the statistical error to quantify the difference between exponents more carefully. With a standard error of $0.02$ for the estimate $\delta = 0.26$, we have that the $1d$ DP value of $\delta = 0.16$ lies five standard errors from the QCP estimate, while the $2d$ DP value of $\delta = 0.45$ lies $9.5$ standard errors away. As such, it seems that the QCP universality class is genuinely different to directed percolation, though the presence of finite-time errors prevents us from stating this with absolute certainty.

Given these results, it is of interest to understand whether the QCP can be associated to some other known universality class and to identify exactly what the relevant quantum contributions might be that push QCP away from  1$d$ DP. An interesting remark in this regard can be made concerning the rapidity reversal symmetry present in DP, \cite{Henkel2008}, which leads to the relation $\alpha = \delta$ between the two exponents characterising the decay of density, see Table \ref{table:exponents}. While we have only investigated the value of $\delta$ in this work, the value of $\alpha$ has been estimated previously in \cite{Carollo2019} to give $\alpha = 0.36 \pm 0.08$. This value lies $5$ standard errors from our estimated $\delta = 0.26$. This seems to suggest that rapidity reversal is indeed broken in the QCP, though confirmation of this will require a better determination of $\alpha$, with QJMC offering a promising approach given the results presented here.

\section*{Acknowledgement}
We thank H. Weimer and M. Buchhold for useful discussions. The  research leading  to  these  results  has  received  funding  from  the European Research Council under the European Unions Seventh  Framework  Programme  (FP/2007-2013)/ERC [grant agreement number 335266 (ESCQUMA)], the Engineering and Physical Sciences Council [grant numbers EP/M014266/1, EP/N03404X/1, EP/R04421X/1], and  the  Leverhulme Trust  [grant  number  RPG-2018-181]. IL gratefully acknowledges funding through the Royal Society Wolfson Research Merit Award. We acknowledge the use of Athena at HPC Midlands+, which was funded by the EPSRC on grant EP/P020232/1, in this research, via the EPSRC RAP call of December 2018.  We are grateful for access to the University of Nottingham's Augusta HPC service.

\bibliographystyle{apsrev4-1}
\bibliography{QCP_vs_CCP.bbl}

\end{document}